# Tailoring of Grain Boundary Structure and Chemistry of Cathode Particles for Enhanced Cycle Stability of Lithium Ion Battery

Pengfei Yan,[a,#] Jianming Zheng,[b,#] Jian Liu,[c] Biqiong Wang [c], Xueliang Sun,[c,*] Chongmin Wang,[a,*] and Ji-Guang Zhang [b,*]

[a] Environmental Molecular Sciences Laboratory, Pacific Northwest National Laboratory, 902 Battelle Boulevard, Richland, WA 99354, USA.
[b] Energy and Environment Directorate, Pacific Northwest National Laboratory, 902 Battelle Boulevard, Richland, WA 99354, USA.
[c] Nanomaterials and Energy Lab, Department of Mechanical and Materials Engineering, University of Western Ontario, London, ON N6A 5B9, Canada

*Corresponding authors: Chongmin.wang@pnnl.gov, xsun9@uwo.ca, Jiguang.zhang@pnnl.gov
# These authors contribute equally to this work.

**ABSTRACT:**

**The biggest challenge for the commercialization of layered structured nickel rich lithium transition metal oxide cathode is the capacity and voltage fading. Resolving this problem over the years follows an incremental progress. In this work, we report our finding of totally a new approach to revolutionize the cycle stability of aggregated cathode particles for lithium ion battery at both room and elevated temperatures. We discover that infusion of a solid electrolyte into the grain boundaries of the cathode secondary particles can dramatically enhance the capacity retention and voltage stability of the battery. We find that the solid electrolyte infused in the boundaries not only acts as a fast channel for Li ion transport, but also most importantly prevents penetration of the liquid electrolyte into the boundaries, consequently eliminating the detrimental factors that include solid-liquid interfacial reaction, intergranular cracking, and layer to spinel phase transformation. The present work, for the first time, reveals unprecedented insight as how the cathode behaves in the case of not contacting with the liquid electrolyte, ultimately points toward a general new route, via grain boundary engineering, for designing of better batteries of both solid-liquid and solid state systems.**

One of the invariable themes for rechargeable lithium-ion batteries (LIB) is to build a better battery with high energy density, long cycle stability, high rate, and safe operation[1, 2, 3, 4]. These objectives are achievable through either exploring new battery materials or optimizing the existing battery components[1, 5, 6, 7, 8, 9, 10]. To facilitate electron and ion transport in electrode, the active particles need to be well dispersed to ensure each particle is in close contact with the conducting additive and wetted by the liquid electrolyte. However, this designing principle intuitively counters against the high density loading of active components and simultaneously leads to the amplification of the deleterious solid-liquid electrolyte reaction, consequently limiting the attainable capacity density and cycle stability of the battery. In order to increase the packing density of cathode, the nano-sized primary particles are intentionally aggregated to form micrometer sized secondary particles, which, however, introduces new challenges[10]. Typically, it has been observed that upon battery cycling, high density intergranular cracks are initiated within secondary particles, leading to the disintegration of the particles and poor cycle stability and eventual failure of the battery[11]. Further, permeation and penetration of the liquid electrolyte along the grain boundaries and cracks in the secondary particles self-amplify the problems associated with solid-liquid reaction[12, 13, 14], as well as the layer to spinel phase transformation, which consequently contributes to the voltage decaying. Advancing of lithium ion battery critically relies on resolving of these bottleneck problems[15, 16].

Tailoring of grain boundary structure and chemistry for optimization of materials behaviors and properties appears to be a classic protocol in material science.[17, 18, 19, 20] However, this strategy has ever hardly been used for the case of rechargeable battery. Herein, we report a new approach, based on tailoring grain boundary structure and chemistry in the secondary particles, to tackle these critical barriers in the system of Ni-rich layered cathode material, $LiNi_{0.76}Mn_{0.14}Co_{0.10}O_2$ (here after referred as Ni-rich NMC) as a specific example. We used a solid electrolyte, $Li_3PO_4$ (here after referred as LPO) to infuse into the grain boundaries of Ni-rich NMC and we found dramatically enhanced cycling stability of the cathode. Detailed structural and chemical analysis combined with electrochemical testing indicate that the thin layer of solid electrolyte in the grain boundary not only can prevent the cracking of the secondary particles and the layer to spinel phase transformation, but also can further change the cathode/electrolyte interfacial kinetics, enabling excellent cycle stability of the Ni-rich cathode.

## Enhanced capacity retention and voltage stability

The Ni-rich NMC is synthesized by co-precipitation method, and the calcination temperature has firstly been carefully optimized to achieve the optimal electrochemical performances (Supplementary Fig. 1). The secondary particles of Ni-rich NMC prepared at optimal condition (750 °C) are further coated with $Li_3PO_4$ (LPO) using atomic layer deposition (ALD).[21, 22] Following the ALD coating, the LPO coated particles are further annealed at 600 °C for 2 hrs in air to enable the infusion of LPO along the grain boundaries of the secondary particles. To reveal the function of LPO, three types of electrodes are fabricated using uncoated, LPO-as-coated, and LPO-infused particles of Ni-rich NMC, respectively. The electrodes are charged/discharged for 200 cycles within a voltage window of 2.7~4.5 V as described in details in the Supplementary Information. The LPO-infused electrode demonstrates the highest capacity retention of 91.6% at room temperature and 73.2% at 60 °C, which is contrasted by 79.0% and 58.3% for the case of using uncoated particles (Fig. 1a and 1b). On the other hand, LPO-as-coated but not infused sample exhibits less capacity as compared with the LPO-infused electrode as shown in Fig. 1 a and b. Associated with the stable capacity retention is the dramatically reduced voltage decay for the case of the LPO-infused particles as shown in Fig. 1c and 1d of the charge/discharge voltage profiles. In terms of rate performance, the LPO-infused electrode shows comparable rate with that of the uncoated one, while the LPO-as-coated electrode shows a poor rate performance (Supplementary Fig. 2). These observations indicate that applying the LPO coating on the outer surface of the secondary particle has limited effect on the performance improvement. However, the annealing in air at 600 °C for 2 hrs not only leads to a superior cycle stability but also enables enhanced rate performance.

## Infusion of solid electrolyte into the grain boundaries

The underlying mechanism as how the infusion process functions to significantly enhance the cycling stability of both capacity and voltage is unveiled by a detailed comparison of the structural and chemical features of these three types of particles at both fresh and cycled states at multiscale spatial resolution. First, the spatial distribution of the coating materials before the battery cycling is identified. For the uncycled particles, phosphorus (P) is unique and can be used as a signature element to trace the spatial distribution of the LPO coating layer. Scanning transmission electron microscopes-energy dispersive X-ray spectroscopy (STEM-EDS) mapping of P distribution indicates that following the ALD coating, the LPO is mainly located at the surface of the secondary particle, with a coating layer thickness of tens nanometer as shown in Fig. 2a and 2b. Following the 2 hrs annealing in air at 600 °C, as shown in Fig. 2c and 2d, the P-enriched surface coating layer disappears. Instead, P-enriched regions are found within the secondary particles. High spatial resolution mapping of P distribution indicates that the grain boundaries and the pockets of the triple grain junction are enriched with P (Fig. 2e, f and Supplementary Fig. 3), demonstrating that during the 600 °C annealing, the LPO coating layer that is initially located at the surface of the secondary particle infuses into the grain boundaries of the secondary particles as schematically illustrated in Fig. 2g though the nominal melting point of $Li_3PO_4$ is 837 °C. The coating and subsequent annealing lead to a slightly increased Li and transition metal interlayer mixing at the outmost surface layer in the LPO-infused electrode as compared with the pristine one (Supplementary Fig. 4). However, the ALD coating and subsequent annealing do not modify the general morphology of the secondary particles (Supplementary Fig. 5).

**Elimination of detrimental factors: solid-liquid interfacial side reaction, intergranular cracking, and layer to spinel phase trnsformation**

After 200 cycles, these three types of particles show distinctively different structural features as representatively demonstrated by the cross-sectional SEM images of the secondary parties. The uncoated particle is featured by the formation of intensive intergranular cracks (Fig. 3a and Supplementary Fig. 6b), which has been identified to be one of the major structural degradation mechanisms that leads to the cathode failure during battery cycling[11, 15, 23]. However, surprisingly, these cracking features do not happen for the LPO-infused samples (Fig. 3e and Supplementary Fig. 6d). Apparently, the infusion of LPO into the grain boundaries of the secondary particles prevents the formation of intergranular cracking in the secondary particles and correspondingly the associated degradation route.

Consistent with the structural difference described above is the chemical differences among these three samples following 200 cycles, reflecting a distinctively different way of interaction of the liquid electrolyte with cathode following the LPO coating and infusion into the grain boundaries. Florine (F) and carbon (C) are the components of the liquid electrolyte but they are not part of cathode particles. Therefore, the spatial distribution of F and C can be used to trace the interaction of the electrolyte and the cathode particles. As shown in Fig. 3b-d, for the uncoated particles, both F and C show enrichment within the cycled secondary particles. This means that the liquid electrolyte penetrates along the grain bounaries of the uncoated secondary particles. The side reaction between the cathode and liquid electrolyte leads to the formation of species such as $Li_2CO_3$, LiF, and $LiFPO_x$[12, 14], as supported by the EDS and electron energy loss sepectrascopy (EELS) analysis shown in the Supplementary Fig. 7. Apparently, formation of these species leads to the consumption of electrolyte, depletion of salt in the electrolyte, formation of a thick SEI layer on both surface and along the grain boundary of the cathode particles. These SEI layers modified the cathode/electrolyte interfacial electrochemistry, which are detrimental to the battery stability. Surprisingly, for the LPO-infused particles, following the battery cycling, C and P only appear at the outer surface of the secondary particles, but not within the secondary particles (see Fig. 3f-h), demonstrating that during the battery cycling the liquid electrolyte does not penetrate into the grain boundaries of the secondary particles, thus, no interfacial side reaction products were observed within the secondary particles. As a direct result, for the LPO-infused electrode, the cathode/electrolyte side reactions only occurre at outer surface of the secondary particles, but not at the primary particles within the secondary particles. The structual and chemical differences among these three types of particles described above are also corroborated by the electrochemcal impedance spectroscopy (EIS) measurment. Attributed to the poor ionic/electronic conductivity induced by the particle cracking and the solid-liquid side reaction, the EIS shows that the charge-transfer resistance across the electrode/electrolyte interface increases much faster for the electrode made up of uncoated particles as compared with those of LPO infused particles(the semicircles at high to medium frequence in Fig. 3i-k).

Electron diffraction and high resolution structural imaging reveal more details as how the infusion of LPO along the grain boundaries of the secondary particles to enhance the battery cycling stability. Blocking of liquid electrolyte from penetrating into the grain boundaries within the secondary particles by the infused LPO eliminates the phase transformation from original layered structure to spinel/rock salt phases, which otherwise typically initiates from the particle surface and propagate inward with the progression of the battery cycling. This point is clearly revealed by the images shown in Fig. 4. As a comparison, the microstructural features of the uncoated particle without cycling are shown at the left panel of Fig. 4a, d, g, j, revealing a dense packing of the layered primary particles within the secondary particle. For the uncoated particle as shown at the middle column of Fig. 4 (b, e, h, k), after 200 cycles, significant structural evolution is identified. The selective area electron diffraction (SAED) pattern is featured by the halo rings with intensity fluctuation (Fig. 4 b), indicating poor crystallinity due to severe lattice distortion and defects generation. The bright-field transmission electron microscopy (TEM) image clearly shows the boundaries between primary particles becomes large and is filled with side reaction products (comparing Fig. 4 d with e). As a consequence of the solid-liquid reaction, the primary particles surface is also covered with a thin surface layer as marked by the arrows in Fig. 4h. At atomic level, the layered structure transforms to rock-salt-like structure (comparing Fig. 4 j with k, and Supplementary Fig. 8). In a sharp contrast, the structural evolutions described above for the uncoated particle do not occur for the LPO-infused particles even after 200 cycles, as clearly revealed by the images shown at the right column of Fig. 4c, f, i, and l.

**Mechanism of enhanced battery cycling stability**

Based on our comprehensive microanalysis, the fundamental mechanisms on why the LPO-infused electrode exhibits superior electrochemical performance over the other two electrode becomes clear. First, the active cathode material inside the secondary particle is well preserved during cycling of the LPO-infused electrode. Even after 200 cycles, the interior region shows no noticeable change. Second, the annealing process redistributes the surfaced coated LPO layer along the grain boundary of the whole secondary particle and enables fast Li-ion transport across cathode without side reaction with liquid electrolyte, evidenced by the first cycle EIS shown in Supplementary Fig. 9. Third, filling of $Li_3PO_4$ solid state electrolyte into the intergranular boundaries can enhance Li-ion transfer thus good rate capability. In addition, the alleviated side reactions translate to minimized electrolyte decomposition/degradation during cycling. Apparently, the surface of the large secondary particle is still in contact with the liquid electrolyte, which contributes to the degradation of the cathode. In principle, the larger of the size of the secondary particle, the smaller of this surface related degradation. From a more general perspective, this work demonstrated a new approach that through grain boundary engineering, one can not only mitigate cathode degradation but also boost Li-ion diffusion kinetics through highly conductive grain boundary pathway. Coupled with further dedicated surface modification to mitigate surface degradation, more pronounced improvement on battery performance can be expected.

It should be pointed out that, due to the infusion of the grain boundaries by the solid electrolyte, the microstructural features thereafter observed as shown in Fig. 4 f-l really represents the behavior of Ni-rich NMC cathode in a solid-state battery configuration. Therefore the observation of the absence of the surface initiated layer to spinel/rock-salt structure at the solid state configuration is significant, providing insight as to what is the controlling factor for this widely documented phase transformation. The present observation conclusively indicates that a direct contact of liquid electrolyte with the layered structure cathode plays a key role for initiation of this phase transformation. Furthermore, this work also provides new insight on the formation mechanism of the intergranular cracks in the secondary particle in general, which has been identified to be one of the major structural degradation mechanisms of the secondary particles that lead to the cathode failure during cycling[11, 15, 16, 24, 25]. It is believed that the anisotropic lattice expansion and contraction during charge/discharge cycle causes significant micro strain among the primary particles and gives rise to the intergranular cracks[23, 26, 27]. For each primary particle, the lattice strain is solely determined by the fraction of the Li that has been extracted from the lattice. Infusion of LPO into the grain boundaries of the secondary particles will not be expected to affect the lattice strain of the primary particle. The absence of intergranular cracking in the LPO-infused particles clearly indicate that the previously observed so-called intergranular cracking in the uncoated particle is essentially the consequences of both build-up of micro strain and the dissolution induced mass loss of the materials due to the penetration of the liquid electrolyte into the grain boundaries. This conclusion is supported by the following four observations. First, the density of the intergranular cracks is very high and the separation gaps between the primary particles becomes large following the cycling of the battery (Fig. 3a and b, Fig. 4e and supplementary Fig. 6b), which cannot be solely explained by the strain induced cracking as the volume shrinkage is expected to be less than 2%. Second, the cracks preferentially align along the radius direction, which is likely related to the transport process of dissolved cations in the liquid upon battery cycling. Third, it is worth to point out that in the cycled LPO-as-coated electrode (without infusion), the intergranular cracks are much less and small as compared with those of uncoated particles, essentially indicating that the coating of the outer surface of the secondary particle partially blocks the liquid penetration of the grain boundaires, consquentially alleviating the dissolution process and therefore a decreased degree of intergranular cracking (Supplementary Fig. 6). Fourth, if the crack is solely due to strain effects, we would expect many cracks in the first few cycles, while the fact is that high density cracks are gradually developed during prolonged cycles[11, 15, 16, 25].

## Conclusions

In summary, we have clearly demonstrated that coating and subsequent infusion of LPO solid state electrolyte along the grain boundaries of secondary particles of a Ni-rich NMC layered cathode can significantly enhance its structural and interfacial stability, therefore, leading to the long-term cycle stability of both capacity and voltage. The detailed structural and chemical analysis reveals that the dramatically enhanced performance is associated with grain boundary modification by the solid state electrolyte, providing a fast path for Li ion transport and simultaneously preventing penetration of liquid electrolyte into the boundary, and therefore eliminating the following detrimental factors: the solid-liquid interfacial reaction, intergranular cracking, and the layer to spinel phase transformation, which critically affects the battery cycle stability of both capacity and voltage  The present work provides a new insight as how an solid-liquid interfacial reaction can affect the battery performance, ultimately points towards a new approach for designing advanced electrode materials through grain boundary engineering with enhancing ionic transport kinetics *via* highly conductive grain boundary pathway. This new strategy opens up a new avenue for the rational design of high performance cathode materials for high-energy-density lithium ion batteries and beyond.

## Methods

**Material preparation.** Spherical Ni-rich $Ni_{0.76}Mn_{0.14}Co_{0.10}(OH)_2$ precursors were prepared by a mixed hydroxide co-precipitation method, using a continuously stirred tank reactor (CSTR) under $N_2$ atmosphere. Initially, the CSTR of 5 L capacity was filled with 1.5 L distilled water. Then, an aqueous solution composed of $NiSO_4$, $MnSO_4$, and $CoSO_4$ with a concentration of 2.0 mol $L^{-1}$ was continuously pumped into the CSTR. Meanwhile, a NaOH solution (4.0 mol $L^{-1}$) as the precipitation reagent and a $NH_4OH$ solution as chelating agent (10 mol $L^{-1}$) were separately fed into the CSTR. The pH value (pH = 11.5), stirring speed (1000 rpm), and temperature (50 °C) were carefully controlled during the precipitation reaction process. The precursor was filtered, thoroughly washed with distilled water, and dried overnight at 110 °C. $LiNi_{0.76}Mn_{0.14}Co_{0.10}O_2$ cathode materials were prepared by mixing the $Ni_{0.76}Mn_{0.14}Co_{0.10}(OH)_2$ precursor powder with LiOH, followed by being sintered at different temperatures for 20 h in air. 3 mol% excess Li was used to compensate the evaporation of Li during calcination at high temperatures.

Deposition of Lithium phosphate on NMC powders was performed in a Savannah 100 ALD system (Ultratech/Cambridge Nanotech.) by using lithium tert-butoxide ($LiO^tBu$) and trimethylphosphate (TMPO) as precursors. The source temperatures for $LiO^tBu$ and TMPO were 180 and 75 °C, respectively, and the deposition temperature for lithium phosphate was 300 °C. Before deposition, NMC powders were well dispersed in a stainless steel tray, which was put in the center of reaction chamber. During one ALD cycle, $LiO^tBu$ and TMPO with a pulse time of 2s was alternatively introduced into the reaction chamber, and pulsing of each precursor was separated by a 15s purge with nitrogen gas. Lithium phosphate with ~ 10 nm thickness (calculated from its growth rate of ~ 0.07 nm/cycle) was coated on the NMC powders by repeating the above ALD cycle for 150 times. The ALD-coated sample was further annealed at 600 °C for 2 h to obtain the ALD-coated-annealed sample.

**Electrochemical measurements.** Electrochemical performance measurements were conducted in R2032 coin-type cells. The thin electrodes were prepared by casting a slurry containing 80% active material, 10% polyvinylidene fluoride binder (PVDF, Kureha L#1120), and 10% super P onto an Al current collector foil. A typical loading of the electrodes is about 4 - 5 mg $cm^{-2}$. After drying, the electrodes were punched into disks with area of 1.27 $cm^2$. Electrochemical cells were assembled with the cathodes as prepared, metallic lithium foil as anode electrode, Cellgard 2500 as separator, and 1 M LiPF$_6$ dissolved in ethyl carbonate (EC) and dimethyl carbonate (DMC) (1:2 in volume) as electrolyte in an argon-filled glove box (Mbraun, Germany) in which both oxygen and moisture content were controlled below 1 ppm. Charge-discharge experiments were performed galvanostatically between 2.7 – 4.5 V on an Arbin BT-2000 battery tester at room temperature and 60 ºC. The rate capability was evaluated using same charge rate of C/5 and a gradual ascending in the discharge C rate after initial five charge/discharge cycles at C/10 rate. A 1C rate corresponds to a current density of 200 mA g$^{-1}$ in this work. Electrochemical impedance spectra (EIS) measurements were performed using a Solartron 1255B frequency analyzer and 1287 electrochemical workstation in a frequency range from 100 kHz to 1 mHz with a perturbation amplitude of ±10 mV.

**Microstructure characterizations.** FIB/SEM imaging and TEM specimen preparation were conducted on FEI Helios DualBeam FIB operating at 2-30 kV. For thin section TEM specimens, they are prepared directly from each electrode foil by a standard lift-out procedure. Firstly, 1.2 μm thick Pt layer (200 nm e-beam deposition followed by 1 μm ion beam deposition) was deposited on a region to void Ga ion beam damage in the subsequent lift-out and thinning process. After lift-out, the specimen was thinned to 200nm using 30kV Ga ion beam. Then, a 2kV final polishing was performed to remove surface damage layer till electron transparency at 5kV SEM imaging. After 2kV Ga ion polish, the surface damage layer is believed to be less than 1 nm.

The FIB-prepared LTMO samples were investigated by FEI Titan80-300 S/TEM microscope at 300 kV. This microscope is equipped with a probe spherical aberration corrector, enabling sub-angstrom imaging using STEM-HAADF detectors. For STEM-HAADF imaging, the inner and outer collection angles of annular dark field detector were set at 55 and 220 mrad, respectively. STEM-EDS and STEM-EELS was performed on a probe aberration-corrected JEOL JEM-ARM200CF at 200 kV. The STEM-EELS data were collected in dual-EESL mode to obtain both zero-loss spectra and core-loss spectra. Core-loss EELS are calibrated by corresponding zero-loss EELS before further analysis using DigitalMicrograph (Version 2.11, Gatan Inc.).

**Data Availability.** All relevant data are kept at the Environmental Molecular Sciences Laboratory storage at Pacific Northwest National Laboratory and are available from the authors on request

**REFERENCES**

1. Lu J, Chen Z, Ma Z, Pan F, Curtiss LA, Amine K. The role of nanotechnology in the development of battery materials for electric vehicles. *Nat Nano* **11**, 1031-1038 (2016).

2. Whittingham MS. Lithium batteries and cathode materials. *Chemical reviews* **104**, 4271-4301 (2004).

3. Etacheri V, Marom R, Elazari R, Salitra G, Aurbach D. Challenges in the development of advanced Li-ion batteries: a review. *Energy & Environmental Science* **4**, 3243-3262 (2011).

4. Tarascon JM, Armand M. Issues and challenges facing rechargeable lithium batteries. *Nature* **414**, 359–367 (2001).

5. Lin D, Liu Y, Cui Y. Reviving the lithium metal anode for high-energy batteries. *Nat Nano* **12**, 194-206 (2017).

6. Manthiram A, Knight JC, Myung ST, Oh SM, Sun YK. Nickel-Rich and Lithium-Rich Layered Oxide Cathodes: Progress and Perspectives. *Adv Energy Mater* **6**, 1501010 (2016).

7. Thackeray MM, Johnson CS, Vaughey JT, Li N, Hackney SA. Advances in manganese-oxide composite electrodes for lithium-ion batteries. *J Mater Chem* **15**, 2257–2267 (2005).

8. Xia H, Meng YS, Lu L, Ceder G. Electrochemical Properties of Nonstoichiometric $LiNi_{0.5}Mn_{1.5}O_{4-\delta}$ Thin-Film Electrodes Prepared by Pulsed Laser Deposition. *J Electrochem Soc* **154**, A737–A743 (2007).

9. Deng H, Belharouak I, Sun Y-K, Amine K. $Li_xNi_{0.25}Mn_{0.75}O_y$ ($0.5 \leqslant x \leqslant 2$, $2 \leqslant y \leqslant 2.75$) compounds for high-energy lithium-ion batteries. *J Mater Chem* **19**, 4510–4516 (2009).

10. Sun Y-K, Myung S-T, Park B-C, Prakash J, Belharouak I, Amine K. High-energy cathode material for long-life and safe lithium batteries. *Nat Mater* **8**, 320-324 (2009).

11. Liu H, *et al.* Intergranular Cracking as a Major Cause of Long-Term Capacity Fading of Layered Cathodes. *Nano Letters*, (2017).

12. Yang P, *et al.* Phosphorus Enrichment as a New Composition in the Solid Electrolyte Interphase of High-Voltage Cathodes and Its Effects on Battery Cycling. *Chemistry of Materials* **27**, 7447-7451 (2015).

13. Edström K, Gustafsson T, Thomas JO. The cathode–electrolyte interface in the Li-ion battery. *Electrochimica Acta* **50**, 397-403 (2004).

14. Xu K. Electrolytes and Interphases in Li-Ion Batteries and Beyond. *Chemical reviews* **114**, 11503-11618 (2014).

15. Lee E-J, *et al.* Development of Microstrain in Aged Lithium Transition Metal Oxides. *Nano Letters* **14**, 4873-4880 (2014).

16. Kim H, Kim MG, Jeong HY, Nam H, Cho J. A New Coating Method for Alleviating Surface Degradation of LiNi0.6Co0.2Mn0.2O2 Cathode Material: Nanoscale Surface Treatment of Primary Particles. *Nano Letters* **15**, 2111-2119 (2015).

17. Luo J, Cheng HK, Asl KM, Kiely CJ, Harmer MP. The Role of a Bilayer Interfacial Phase on Liquid Metal Embrittlement. *Science* **333**, 1730-1733 (2011).

18. Cho J, Wang CM, Chan HM, Rickman JM, Harmer MP. Role of segregating dopants on the improved creep resistance of aluminum oxide. *Acta Materialia* **47**, 4197-4207 (1999).

19. Buban JP, *et al.* Grain Boundary Strengthening in Alumina by Rare Earth Impurities. *Science* **311**, 212-215 (2006).

20. Shibata N, Pennycook SJ, Gosnell TR, Painter GS, Shelton WA, Becher PF. Observation of rare-earth segregation in silicon nitride ceramics at subnanometre dimensions. *Nature* **428**, 730-733 (2004).

21. Li X, *et al.* Atomic layer deposition of solid-state electrolyte coated cathode materials with superior high-voltage cycling behavior for lithium ion battery application. *Energy & Environmental Science* **7**, 768-778 (2014).

22. Jian L, Xueliang S. Elegant design of electrode and electrode/electrolyte interface in lithium-ion batteries by atomic layer deposition. *Nanotechnology* **26**, 024001 (2015).

23. Miller DJ, Proff C, Wen JG, Abraham DP, Bareño J. Observation of Microstructural Evolution in Li Battery Cathode Oxide Particles by In Situ Electron Microscopy. *Adv Energy Mater* **3**, 1098-1103 (2013).

24. Sasaki T, Godbole V, Takeuchi Y, Ukyo Y, Novák P. Morphological and Structural Changes of Mg-Substituted Li(Ni,Co,Al)O2 during Overcharge Reaction. *Journal of The Electrochemical Society* **158**, A1214-A1219 (2011).

25. Yan P, Zheng J, Gu M, Xiao J, Zhang J-G, Wang C-M. Intragranular cracking as a critical barrier for high-voltage usage of layer-structured cathode for lithium-ion batteries. *Nature Communications* **8**, 14101 (2017).

26. Lim J-M, Hwang T, Kim D, Park M-S, Cho K, Cho M. Intrinsic Origins of Crack Generation in Ni-rich LiNi0.8Co0.1Mn0.1O2 Layered Oxide Cathode Material. *Scientific Reports* **7**, 39669 (2017).

27. Wang H, Jang YI, Huang B, Sadoway DR, Chiang YM. TEM Study of Electrochemical Cycling‐Induced Damage and Disorder in LiCoO2 Cathodes for Rechargeable Lithium Batteries. *Journal of The Electrochemical Society* **146**, 473-480 (1999).

 Acknowledgements

This work is supported by the Assistant Secretary for Energy Efficiency and Renewable Energy, Office of Vehicle Technologies of the U. S. Department of Energy under Contract No. DE-AC02-05CH11231, Subcontract No. 18769 and No. 6951379 under the Advanced Battery Materials Research (BMR) program. The microscopic analysis in this work was conducted in the William R. Wiley Environmental Molecular Sciences Laboratory (EMSL), a national scientific user facility sponsored by DOE's Office of Biological and Environmental Research and located at PNNL.  PNNL is operated by Battelle for the Department of Energy under Contract DE-AC05-76RLO1830.  Solid-state electrolyte coating by ALD was conducted in Dr. Sun's lab and is financially supported by Nature Sciences and Engineering Research Council of Canada (NSERC) Program, Canada Research Chair (CRC) Program, Canada Foundation for Innovation (CFI), and University of Western Ontario.  A US patent application (US 31063-E) including the grain boundary engineering in cathode partially based on the concept of this work has been filed.

**Author contributions**

C.W., J.Z., and J.-G.Z. initiated this research project. J.Z. and J.L. synthesized cathode materials. J. L. B. W. and X. S. carried out the ALD coating, J.Z. performed battery test. P.Y. conducted the TEM analysis. P.Y., J.Z., C.W. and J.-G.Z. prepared the manuscript with the input from all other co-authors.

**Additional information**

Supplementary Information is available for this paper

**Competing interests**

The authors declare no competing financial interests

# Figure captions

**Figure 1 | Effect of Li₃PO₄ infusion on the electrochemical performance.** Cycling performance of the three $LiNi_{0.76}Mn_{0.14}Co_{0.10}O_2$ cathodes after 3 formation cycles at the voltage range of 2.7 – 4.5 V at **a**, room temperature and **b**, 60 °C. The charge/discharge rate is C/3 at room temperature and C/2 at 60 °C. The corresponding charge/discharge voltage profile evolution of **c**, pristine and **d**, Li₃PO₄-infused $LiNi_{0.76}Mn_{0.14}Co_{0.10}O_2$ cathodes at room temperature.

**Figure 2 | Tracking the spatial distribution of Li₃PO₄ through mapping P prior to battery cycling. a,b**, STEM image and EDS map of P in the as-coated sample, indicating Li₃PO₄ as a layer covering the surface of the secondary particle. The scale bar is 500 nm. **c,d**, STEM image and EDS map of P the **Li₃PO₄**-infused sample, indicating infusion of Li₃PO₄ in the secondary particles. The scale bar is 1 μm. **e,f**, a high resolution STEM image and EDS map of P from a region in **c**, indicating the penetration of Li₃PO₄ along the grain boundaries in the secondary particles. The scale bar is 20 nm. **g**, Schematic illustration showing the evolution of the Li₃PO₄ coating layer on the secondary particle following the coating and annealing.

**Figure 3 | Infusion of Li₃PO₄ into secondary particles eliminates intergranular cracking. a**, Cross sectional SEM image and **b**, STEM-HAADF image and **c,d**, corresponding C and F maps (two signature elements from the electrolyte) from the pristine electrode after 200 cycles, indicating the formation of intergranular cracks and penetration of electrolyte along the grain boundaries. **e**, Cross sectional SEM image, and **f**, STEM-HAADF image and **g,h**, corresponding C and F maps from the Li₃PO₄-infused electrode after 200 cycles, showing no intergranular cracks and absence of electrolyte related species along the grain boundaries in the secondary particles. The scale bars are 2 μm. **i-k**, Impedance spectra evolution of the pristine electrode and the Li₃PO₄-infused electrode at $1^{st}$, $10^{th}$ and $50^{th}$ cycles.

**Figure 4 | Infusion of Li₃PO₄ into secondary particles eliminates structural degradation.** The structural degradations are evaluated by a combination of selected area electron diffraction **a-c**, bright-field TEM imaging **d-f**, STEM-HAADF imaging **g-i** and atomic level STEM-HAADF imaging **j-l** (corresponding to the high magnification image of the red line marked regions). The left column corresponds to the pristine electrode without cycling **a,d,g,j**, the middle column is the pristine electrode after 200 cycles, and the right column shows the Li₃PO₄-infused electrode after 200 cycles. These observations demonstrate that following 200 cycles, the pristine electrode shows significant structural degradation as indicated by comparing the middle column with left column, featuring intergranular cracking and formation of amorphous phase within the grain boundaries (see **e**), formation of surface reaction layer on each grain surface (indicated in **h** by the yellow arrows), and layer to spinel transformation (see **k**), while these degradation features do not happen in the Li₃PO₄-infused electrode (compare right column with left and middle columns). The scale bars are 500 nm in **d-f**, 100 nm in **g-i** and 2 nm in **j-l**, respectively.

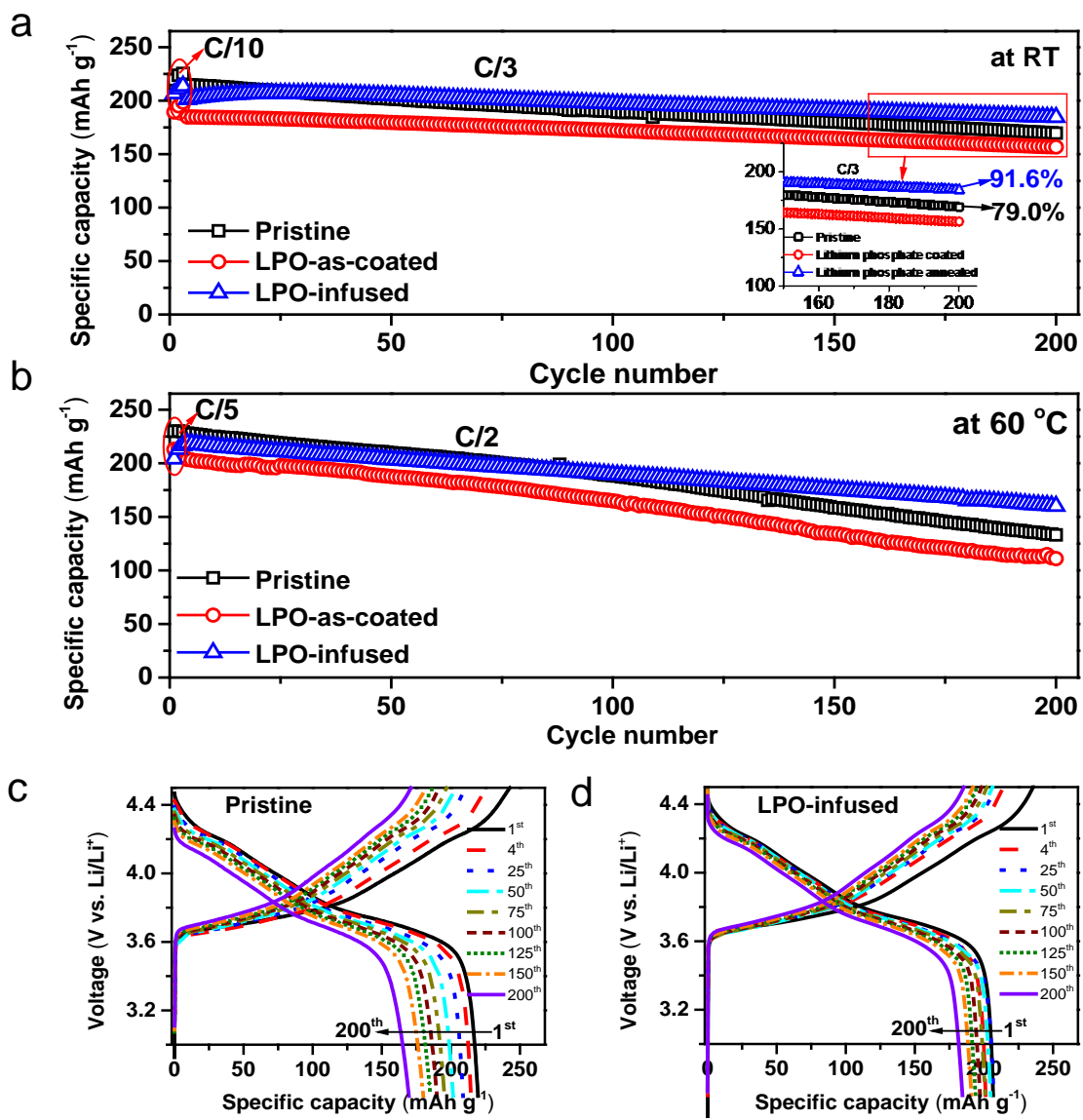

**Figure 1 | Effect of Li$_3$PO$_4$ infusion on the electrochemical performance.** Cycling performance of the three LiNi$_{0.76}$Mn$_{0.14}$Co$_{0.10}$O$_2$ cathodes after 3 formation cycles at the voltage range of 2.7 – 4.5 V at **a**, room temperature and **b**, 60 °C. The charge/discharge rate is C/3 at room temperature and C/2 at 60 °C. The corresponding charge/discharge voltage profile evolution of **c**, pristine and **d**, Li$_3$PO$_4$-infused LiNi$_{0.76}$Mn$_{0.14}$Co$_{0.10}$O$_2$ cathodes at room temperature.

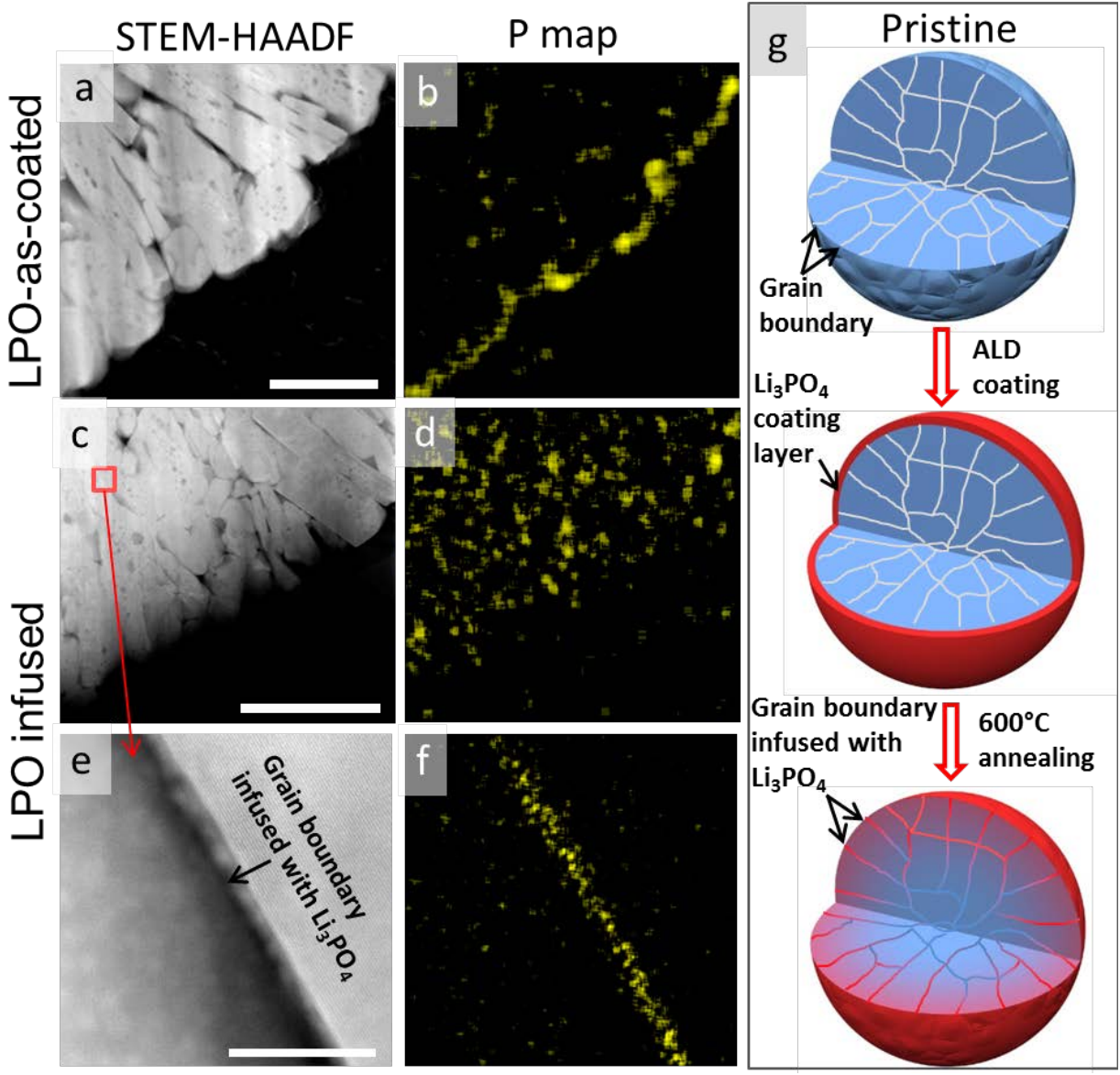

**Figure 2 | Tracking the spatial distribution of Li$_3$PO$_4$ through mapping P prior to battery cycling. a,b**, STEM image and EDS map of P in the as-coated sample, indicating Li$_3$PO$_4$ as a layer covering the surface of the secondary particle. The scale bar is 500 nm. **c,d**, STEM image and EDS map of P the Li$_3$PO$_4$-infused sample, indicating infusion of Li$_3$PO$_4$ in the secondary particles. The scale bar is 1 μm. **e,f**, a high resolution STEM image and EDS map of P from a region in **c**, indicating the penetration of Li$_3$PO$_4$ along the grain boundaries in the secondary particles. The scale bar is 20 nm. **g**, Schematic illustration showing the evolution of the Li$_3$PO$_4$ coating layer on the secondary particle following the coating and annealing.

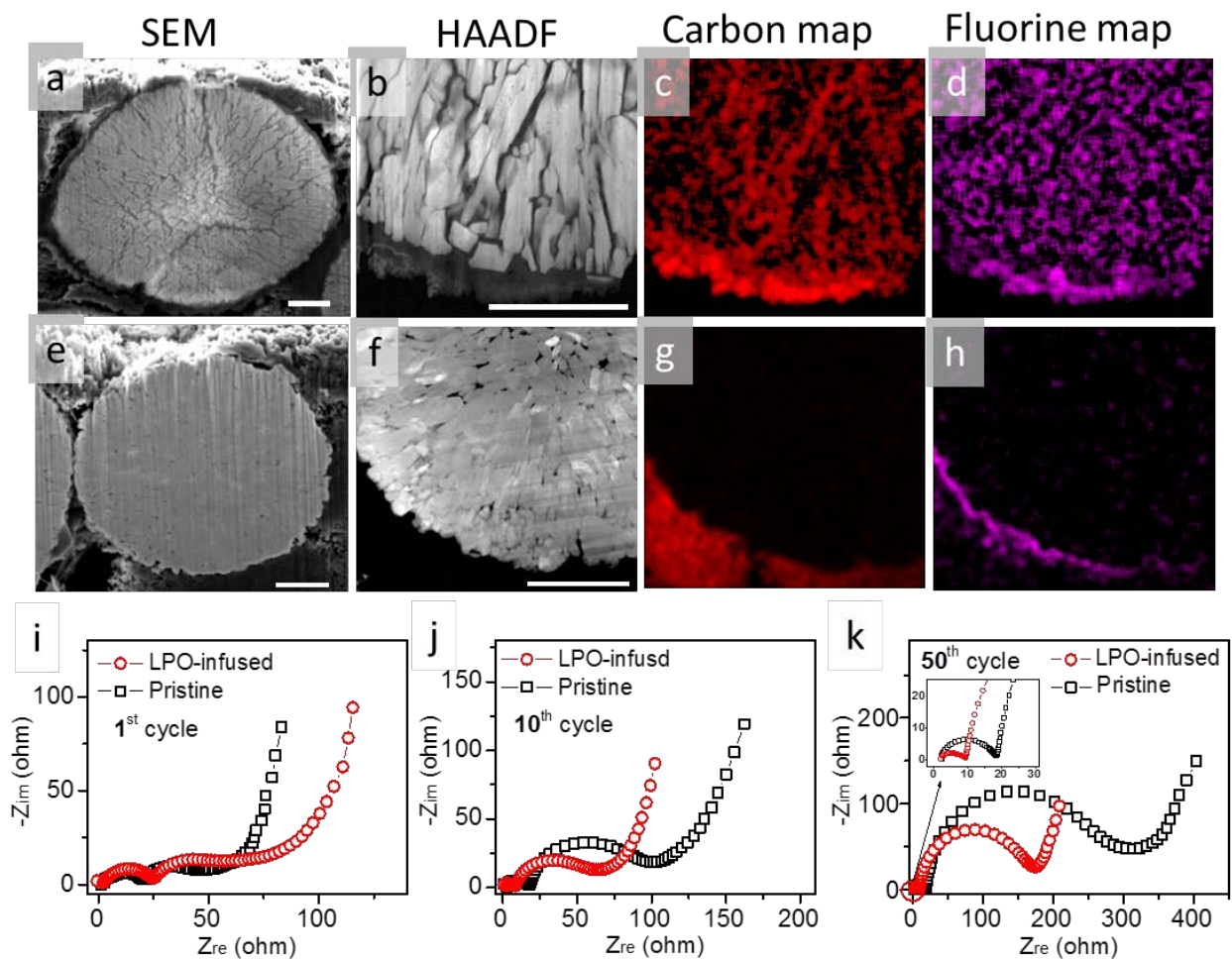

**Figure 3 | Infusion of Li₃PO₄ into secondary particles eliminates intergranular cracking. a**, Cross sectional SEM image and **b**, STEM-HAADF image and **c,d**, corresponding C and F maps (two signature elements from the electrolyte) from the pristine electrode after 200 cycles, indicating the formation of intergranular cracks and penetration of electrolyte along the grain boundaries. **e** Cross sectional SEM image, and **f**, STEM-HAADF image and **g,h**, corresponding C and F maps from the Li₃PO₄-infused electrode after 200 cycles, showing no intergranular cracks and absence of electrolyte related species along the grain boundaries in the secondary particles. The scale bars are 2 μm. **i-k**, Impedance spectra evolution of the pristine electrode and the Li₃PO₄-infused electrode at 1$^{st}$, 10$^{th}$ and 50$^{th}$ cycles.

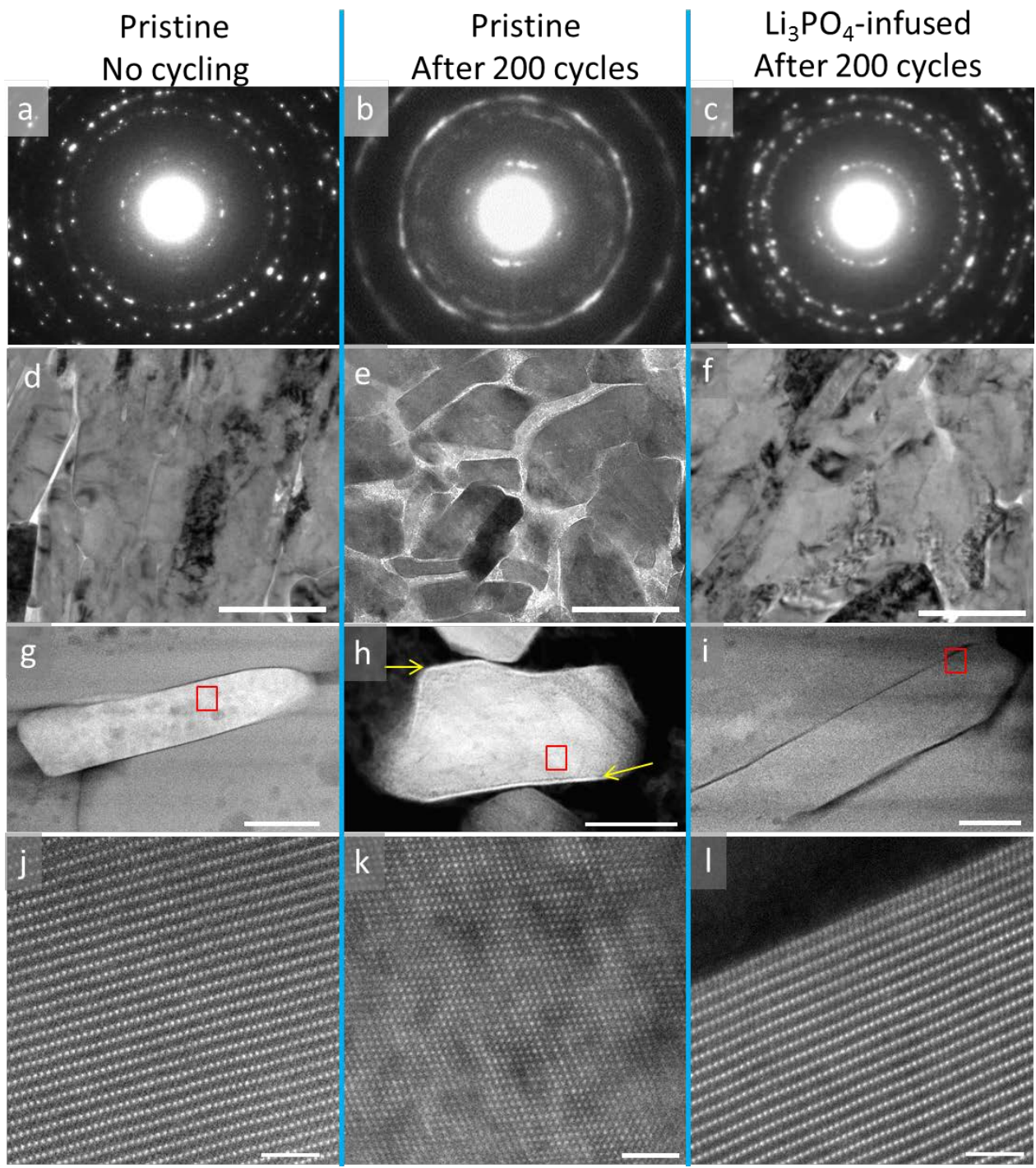

**Figure 4 | Infusion of Li₃PO₄ into secondary particles eliminates structural degradation.** The structural degradations are evaluated by a combination of selected area electron diffraction **a**-**c**, bright-field TEM imaging **d**-**f**, STEM-HAADF imaging **g**-**i** and atomic level STEM-HAADF imaging **j**-**l** (corresponding to the high magnification image of the red line marked regions). The left column corresponds to the pristine electrode without cycling **a**,**d**,**g**,**j**, the middle column is the pristine electrode after 200 cycles, and the right column shows the Li₃PO₄-infused electrode after 200 cycles. These observations demonstrate that following 200 cycles, the pristine electrode shows
21

significant structural degradation as indicated by comparing the middle column with left column, featuring intergranular cracking and formation of amorphous phase within the grain boundaries (see **e**), formation of surface reaction layer on each grain surface (indicated in **h** by the yellow arrows), and layer to spinel transformation (see **k**), while these degradation features do not happen in the $Li_3PO_4$-infused electrode (compare right column with left and middle columns). The scale bars are 500 nm in **d**-**f**, 100 nm in **g**-**i** and 2 nm in **j**-**l**, respectively.